\documentclass{article}
\usepackage{spconf,amsmath,graphicx,hyperref}
\usepackage{cite}
\usepackage{multirow, multicol}
\usepackage{makecell}
\usepackage{bbding}
\usepackage{enumitem}
\PassOptionsToPackage{hyphens}{url}
\usepackage{url}
\usepackage{hyperref}
\usepackage{amssymb}
\usepackage{booktabs}

\title{DAIEN-TTS: Disentangled Audio Infilling for Environment-Aware Text-to-Speech Synthesis}
\name{Ye-Xin Lu$^1$, Yu Gu, Kun Wei, Hui-Peng Du$^1$, Yang Ai\sthanks{Corresponding author. This work was supported in part by the National Nature Science Foundation of China under Grant U23B2053 and 62301521.}$^1$, Zhen-Hua Ling$^1$}
\address{$^1$National Engineering Research Center of Speech and Language Information Processing, \\University of Science and Technology of China, Hefei, P. R. China \\
{\small \tt \ \{yxlu0102, redmist\}@mail.ustc.edu.cn, \{yangai, zhling\}@ustc.edu.cn}}

\begin{document}
\ninept
\maketitle
\vspace{-1mm}
\begin{abstract}
\vspace{-1mm}
This paper presents DAIEN-TTS, a zero-shot text-to-speech (TTS) framework that enables ENvironment-aware synthesis through Disentangled Audio Infilling.
By leveraging separate speaker and environment prompts, DAIEN-TTS allows independent control over the timbre and background environment of the synthesized speech.
Built upon F5-TTS, the proposed DAIEN-TTS first incorporates a pretrained speech-environment separation (SES) module to disentangle the environmental speech into mel-spectrograms of clean speech and environment audio.
Two random span masks of varying lengths are then applied to both mel-spectrograms, which, together with the text embedding, serve as conditions for infilling the masked environmental mel-spectrogram, enabling the simultaneous continuation of personalized speech and time-varying environmental audio. 
To further enhance controllability during inference, we adopt dual classifier-free guidance (DCFG) for the speech and environment components and introduce a signal-to-noise ratio (SNR) adaptation strategy to align the synthesized speech with the environment prompt. 
Experimental results demonstrate that DAIEN-TTS generates environmental personalized speech with high naturalness, strong speaker similarity, and high environmental fidelity.

\end{abstract}
\vspace{-1mm}
\section{Introduction}
\vspace{-1mm}
\label{sec:intro}
Zero-shot text-to-speech (TTS) refers to synthesizing speech for unseen speakers conditioned on a short speech prompt from the speaker.
Recently, advanced zero-shot TTS methods have adopted the in-context learning paradigm, such as neural-codec-based language modeling \cite{wang2023neural, wang2024speechx}, flow-matching-based speech infilling \cite{le2023voicebox, chen2024f5}, and their hybrids \cite{anastassiou2024seed, du2024cosyvoice, lee2024ditto}, which have achieved substantial improvements in speech naturalness and speaker similarity.
Typically, zero-shot TTS systems are expected to produce high-quality personalized speech, while in-context-learning-based models tend to preserve the acoustic environment characteristics of the speech prompt, motivating the development of noise-robust zero-shot TTS methods \cite{wang2024investigation, zhang2025advanced, lu2025improving}.
However, in scenarios such as audiobooks and virtual reality, there is often a need to synthesize speech with background environments. 
Although some existing methods \cite{zhang2025advanced, peng2024voicecraft} can effectively preserve the background environment of the speech prompt, they still fail to disentangle the environment for independent control.

Environment-aware TTS refers to the task of using separate speaker and environment prompts to independently control the timbre and acoustic background of the synthesized speech. 
Previous methods \cite{tan2021environment, lu2025incremental} employed dedicated speaker and environment encoders to extract global embeddings for disentangled control.
However, such approaches were constrained to time-invariant acoustic environments such as reverberation and struggled with time-varying background noise.
Recent text-to-audio (TTA) methods have demonstrated the ability to generate environmental audio with speech \cite{lee2024voiceldm, jung2025voicedit}. 
Nevertheless, VoiceLDM \cite{lee2024voiceldm} often produced repetitive and mumbled speech, while VoiceDiT \cite{jung2025voicedit} lacked zero-shot capability and suffered from poor alignment between speech and environment.
UmbraTTS \cite{glazer2025umbratts} enabled environmental personalized speech synthesis using separate clean speech and environment audio prompts, with additional control over the signal-to-noise ratio (SNR) via a speech-to-environment ratio (SER) embedding.
Yet, it assumed access to pure clean and pure environment audio prompts of equal length, which is impractical in real-world scenarios.

In this paper, we propose DAIEN-TTS, an ENvironment-ware zero-shot TTS framework based on Disentangled Audio Infilling, enabling independent control of timbre and background environment via speaker and environment prompts.
Built upon the flow-matching-based F5-TTS framework \cite{chen2024f5}, DAIEN-TTS incorporates a pretrained speech–environment separation (SES) module to extract clean speech and environment mel-spectrograms from environmental speech.
During training, we apply random span masks of varying lengths to simulate diverse prompt durations.
The unmasked environment mel-spectrogram is integrated into the Diffusion Transformer (DiT) block \cite{peebles2023scalable} via cross-attention, alongside the concatenation of the unmasked speech mel-spectrogram and text embedding, enabling the simultaneous infilling of speech and background environment.
During inference, we employ dual classifier-free guidance (DCFG) to enhance the controllability of the speech and environment components and further introduce an SNR adaptation strategy to align the synthesized speech with the environment prompt.
Our main contributions are as follows:

1) We propose DAIEN-TTS, the first environment-aware zero-shot TTS framework capable of disentangling and independently controlling timbre and time-varying background environments.

2) We introduce a cross-attention-based conditioning scheme for effective environment control, and further enhance generation controllability at inference with DCFG and SNR adaptation.

3) Extensive experiments show that DAIEN-TTS achieves high naturalness, speaker similarity, and environmental fidelity across both objective and subjective evaluations.


\begin{figure*}[ht!]
  \centering
  \includegraphics[width=0.95\textwidth]{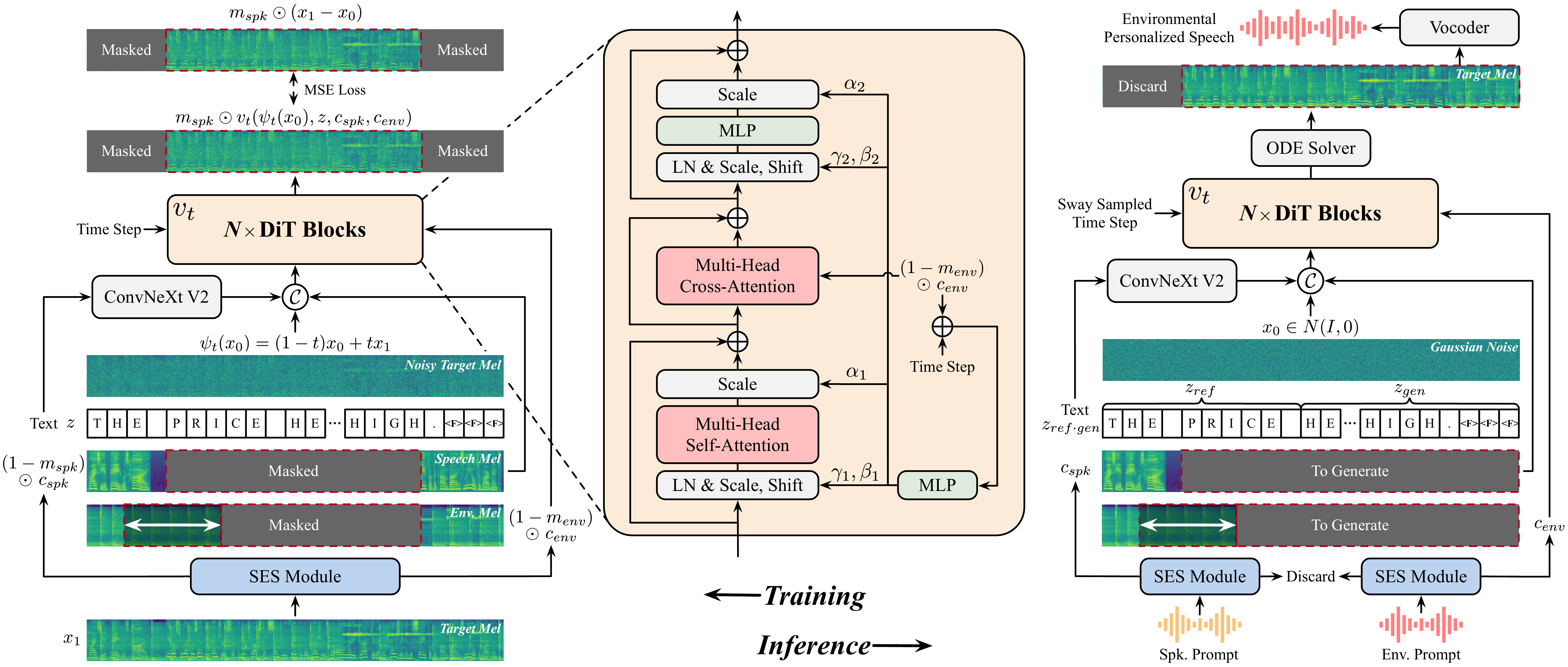}
  \caption{Overview of the proposed DAIEN-TTS training (left) and inference (right) processes.} 
  \label{fig: model}
\end{figure*}

\vspace{-1mm}
\section{Methodology}
\vspace{-1mm}
\label{sec:method}
The overall structure of the proposed DAIEN-TTS is illustrated in Fig.~\ref{fig: model}.
Built upon F5-TTS \cite{chen2024f5}, which formulates zero-shot TTS as a text-guided speech infilling task under a flow-matching framework \cite{le2023voicebox}, DAIEN-TTS proposes a disentangled audio infilling method to enable time-varying environment-aware speech synthesis.
Given an environmental speech signal $\boldsymbol{y} \in \mathbb{R}^{T}$, where $T$ is the waveform sequence length, a pretrained Transformer-based SES module is introduced to decompose $\boldsymbol{y}$ into a clean speech mel-spectrogram $\boldsymbol{c}_{spk} \in \mathbb{R}^{M \times L}$ and an environmental audio mel-spectrogram $\boldsymbol{c}_{env} \in \mathbb{R}^{M \times L}$, where $M$ denotes the mel dimension and $L$ is the spectrogram sequence length.
To simulate varying prompt conditions, random span masks of different lengths $\boldsymbol{m}_{spk}$ and $\boldsymbol{m}_{env}$ (with values of 0 or 1) are applied to $\boldsymbol{c}_{spk}$ and $\boldsymbol{c}_{env}$, respectively.
Finally, the unmasked portions, $(1 - \boldsymbol{m}_{spk}) \odot \boldsymbol{c}_{spk}$ and $(1 - \boldsymbol{m}_{env}) \odot \boldsymbol{c}_{env}$, together with the extended text sequence $\boldsymbol{z}$ with a length of $L$, are used as conditioned inputs to reconstruct the masked environmental speech component $\boldsymbol{m}_{spk} \odot \boldsymbol{x}_1$, where $\boldsymbol{x}_1$ is the mel-spectrogram of the environmental speech $\boldsymbol{y}$.
The details of the model structure, along with the training and inference pipelines, are described as follows.

\begin{figure}[t!]
  \centering
  \includegraphics[width=0.95\linewidth]{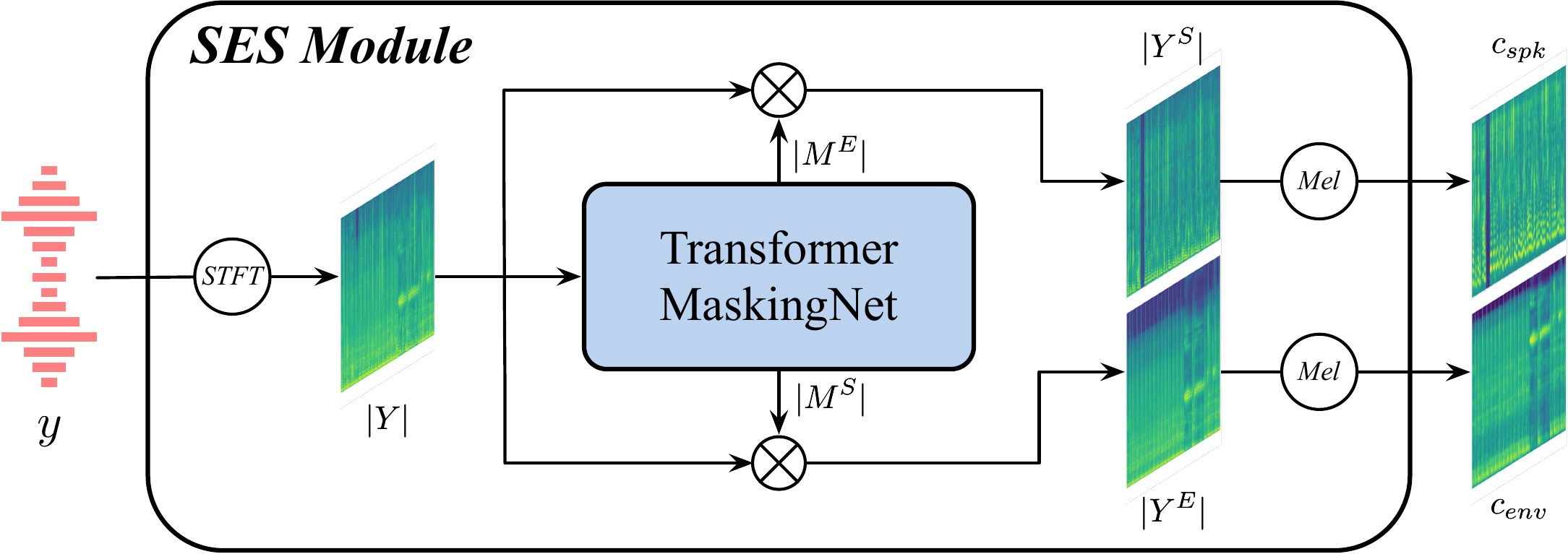}
  \caption{Model structure of the SES module.} 
  \label{fig: ses}
\end{figure}

\vspace{-1mm}
\subsection{SES Module}
\vspace{-1mm}
As illustrated in Fig.~\ref{fig: ses}, the SES module performs speech and background environment separation at the magnitude spectrogram level. 
Specifically, the environmental speech $\boldsymbol{y}$ is first transformed into the magnitude spectrogram $\vert \boldsymbol{Y} \vert \in \mathbb{R}^{F \times L}$ via short-time Fourier transform (STFT), where $F$ denotes the number of frequency bins. 
Subsequently, the environmental magnitude spectrogram $\vert \boldsymbol{Y} \vert$ is fed into a Transformer-based masking net, which consists of an input convolutional layer, $K$ Transformer blocks, and two output convolutional layers.
The masking net predicts two masks $\vert \boldsymbol{M}^{S} \vert$ and $\vert \boldsymbol{M}^{E} \vert \in \mathbb{R}^{F \times L}$, corresponding to the speech and background environment components, respectively.
These masks are applied to $\vert \boldsymbol{Y} \vert$ via element-wise multiplication to suppress the corresponding unwanted components, producing the separated speech and environment magnitude spectrograms  $\vert \boldsymbol{Y}^{S} \vert = \vert \boldsymbol{Y} \vert \odot \vert \boldsymbol{M}^{E} \vert \in \mathbb{R}^{F \times L}$ and $\vert \boldsymbol{Y}^{E} \vert = \vert \boldsymbol{Y} \vert \odot \vert \boldsymbol{M}^{S} \vert \in \mathbb{R}^{F \times L}$, respectively.
Finally, both of them are passed through a mel filter bank to obtain their respective mel-spectrograms, which serve as speaker and environment conditions for the subsequent audio infilling process.

\vspace{-1mm}
\subsection{Training}
\vspace{-1mm}
The training process of the proposed DAIEN-TTS is illustrated in the left part of Fig.~\ref{fig: model}.
Similar to F5-TTS, we train our model on a text-guided audio infilling task, which aims to predict a masked segment of environmental speech, conditioned on the full text (including the transcriptions of the surrounding part and the part to generate), as well as its surrounding clean speech and background environment audio disentangled by the SES module.
To enable the model to handle speaker and environment prompts of varying lengths during inference, we apply random span masks of different lengths to the disentangled speech mel-spectrogram $\boldsymbol{c}_{spk}$ and environment mel-spectrogram $\boldsymbol{c}_{env}$.
Based on the conditional flow matching (CFM) mechanism, we feed the noisy environmental speech $\psi_t(\boldsymbol{x}_0) = (1-t)\boldsymbol{x}_0 + t\boldsymbol{x}_1$ as the model input, where $\boldsymbol{x}_0$ is the sampled Gaussian noise and $t$ is the randomly sampled flow step.
In contrast to UmbraTTS \cite{glazer2025umbratts}, which directly concatenates all conditions with the noisy input to the DiT block, we insert a multi-head cross-attention layer into each DiT block to introduce $(1 - \boldsymbol{m}_{env}) \odot \boldsymbol{c}_{env}$ and inject environment information \cite{lee2024ditto, jung2025voicedit}.
Finally, the model is trained to reconstruct $\boldsymbol{m}_{spk} \odot \boldsymbol{x}_1$ with the conditions of $(1 - \boldsymbol{m}_{spk}) \odot \boldsymbol{c}_{spk}$, $(1 - \boldsymbol{m}_{env}) \odot \boldsymbol{c}_{env}$, and the extended text $\boldsymbol{z}$ using the following training objective:
\begin{align}
\mathcal{L}_{\mathrm{CFM}} = \big \Vert &
\big( v_t(\psi_t(\boldsymbol{x}_0) \vert \boldsymbol{z}, 
(1 - \boldsymbol{m}_{spk}) \odot \boldsymbol{c}_{spk}, 
(1 - \boldsymbol{m}_{env}) \odot \boldsymbol{c}_{env}) \notag \\
& - (\boldsymbol{x}_1 - \boldsymbol{x}_0) \big) 
\odot  \boldsymbol{m}_{spk} \big\Vert_2^2 \label{eq:cfm},
\end{align}
where $v_t(\cdot)$ is the learned velocity field.

\subsection{Inference}
The inference process of the proposed DAIEN-TTS is illustrated in the right part of Fig.~\ref{fig: model}.
Given a speaker prompt speech $\boldsymbol{y}_{spk}$ and an environment prompt speech $\boldsymbol{y}_{env}$, we first employ the SES module to separate the speech component from $\boldsymbol{y}_{spk}$ and the background environment component from $\boldsymbol{y}_{env}$, which serve as the speaker condition $\boldsymbol{c}_{spk}$ and the environment condition $\boldsymbol{c}_{env}$, respectively.
With the transcription of the speaker prompt $\boldsymbol{z}_{ref}$ and a text prompt $\boldsymbol{z}_{gen}$ for generation, we concatenate them together as $\boldsymbol{z}_{ref \cdot gen}$ and use the ordinary differential equation (ODE) solver to generate the target mel-spectrogram with the desired content.
Starting from a sampled noise $\psi_0(\boldsymbol{x}_0) = \boldsymbol{x}_0$, the ODE solver integrates towards $\psi_1(\boldsymbol{x}_0) = \boldsymbol{x}_1$ according to the differential equation:
\begin{equation}
d \psi_t(\boldsymbol{x}_0)/dt = v_t(\psi_t(\boldsymbol{x}_0), \boldsymbol{z}_{ref \cdot gen}, \boldsymbol{c}_{spk}, \boldsymbol{c}_{env}).
\end{equation}
After obtaining the generated target mel-spectrogram, we discard the $\boldsymbol{c}_{spk}$ portion and use a vocoder to reconstruct the environmental personalized speech.

\subsubsection{Dual Classifier-Free Guidance}
F5-TTS employs CFG \cite{ho2022classifier} to achieve a trade-off between generation fidelity and diversity.
Although DAIEN-TTS is conditioned on three items, the text and speaker information are jointly embedded in the speech signal and independent of the background environment.
Therefore, we treat them as a unified guidance term during inference.
Based on this, we adopt a DCFG mechanism \cite{lee2024voiceldm} to enhance the controllability of the speech and background environment component of the synthesized environmental speech:
\begin{align}
        v_{t, \mathrm{DCFG}} = & v_t(\psi_t(\boldsymbol{x}_0), \boldsymbol{z}, \boldsymbol{c}_{spk}, \boldsymbol{c}_{env}) \notag \\ +  \alpha_{speech}&\big( v_t(\psi_t(\boldsymbol{x}_0), \boldsymbol{z}, \boldsymbol{c}_{spk}, \emptyset) - v_t(\psi_t(\boldsymbol{x}_0), \emptyset, \emptyset, \emptyset)\big) \notag \\ +  \alpha_{env}&\big( v_t(\psi_t(\boldsymbol{x}_0), \emptyset, \emptyset, \boldsymbol{c}_{env}) - v_t(\psi_t(\boldsymbol{x}_0), \emptyset, \emptyset, \emptyset)\big),
\end{align}
where $\emptyset$ indicates the null condition, $\alpha_{speech}$ and $\alpha_{env}$ are the DCFG strength for the speech and environment conditions.

\subsubsection{SNR Adaptation}
For the environment prompt, we aim not only to imitate its background acoustic environment but also to preserve its SNR in the synthesized speech.
To achieve this, we introduce an SNR adaptation strategy during inference.
Considering the environmental speech $\boldsymbol{y}$ is composed of its speech component $\boldsymbol{y}^S$ and background environment component $\boldsymbol{y}^E$, such that $\boldsymbol{y} = \boldsymbol{y}^S + \boldsymbol{y}^E$.
According to the definition of SNR and the discrete Parseval’s theorem, we have:
\begin{equation}
    SNR = 10 \log \bigg(\frac{\Vert \boldsymbol{y}^S \Vert_2^2}{\Vert \boldsymbol{y}^E \Vert_2^2}\bigg) = 10 \log \bigg(\frac{\Vert \boldsymbol{Y}^S \Vert_2^2}{\Vert \boldsymbol{Y}^E \Vert_2^2}\bigg).
\end{equation}
Therefore, we can apply a scaling factor to the separated background environment magnitude spectrogram $\boldsymbol{Y}^E_{env}$ to ensure the SNR of the synthesized speech matches that of the environment prompt:
\begin{equation}
    10 \log \bigg(\frac{\Vert \boldsymbol{Y}^S_{spk} \Vert_2^2}{\Vert \boldsymbol{Y}^E_{env} \cdot Scale \Vert_2^2}\bigg) = 10 \log \bigg(\frac{\Vert \boldsymbol{Y}^S_{env} \Vert_2^2}{\Vert \boldsymbol{Y}^E_{env}\Vert_2^2}\bigg).
\end{equation}
Accordingly, we can derive that $ Scale = \sqrt{\Vert \boldsymbol{Y}^S_{spk} \Vert_2^2 / \Vert\boldsymbol{Y}^S_{env}\Vert_2^2}$.
During inference, $\boldsymbol{Y}^E_{env}$ is scaled by the calculated factor and then transformed into a mel-spectrogram, which serves as the final environment condition for environmental speech generation.

\vspace{-1mm}
\section{Experiments}
\vspace{-1mm}
\label{sec:exp}
\subsection{Dataset and Experimental Setup}
\vspace{-1mm}
We employed the LibriTTS corpus \cite{zen2019libritts}, which contains 580 hours of speech data, to train the proposed DAIEN-TTS model.
To simulate a wide range of time-varying background environments, the speech data was mixed with environmental audio from the DNS-Challenge dataset \cite{dubey2022icassp}, using SNRs uniformly sampled between -5 dB and 15 dB.
The DNS-Challenge dataset comprises 68,000 environment audio clips sourced from AudioSet \cite{gemmeke2017audio} and FreeSound \footnote{\href{https://freesound.org}{https://freesound.org}.}.
For evaluation, we adopted the SeedTTS \textit{test-en} set \cite{anastassiou2024seed} as the speech test set and augmented it with environment audio clips from SoundBible \footnote{\href{https://soundbible.com}{https://soundbible.com}.} using SNRs uniformly distributed between 0 dB and 20 dB.
All clean-environmental speech pairs were constructed at a sampling rate of 24 kHz.

In the SES module, we set the number of Transformer layers $K$ to 8, with 16 attention heads, an embedding dimension of 1024, and a feed-forward network (FFN) dimension of 2048. 
In the TTS module, each DiT block is augmented with a cross-attention layer, also configured with 16 attention heads.
During training, we adopted a dynamic mixing strategy to construct clean–environmental speech pairs.
For pretraining the SES module, all speech data was mixed with environment audio to ensure robust separation learning.
For TTS training, we applied a probabilistic augmentation strategy, where each speech sample was mixed with environmental audio with a probability of 50\%, and with silent segments otherwise, in order to encourage better alignment learning between text and speech.
Both the SES module and the TTS module were trained for 600k steps with a batch size of 102,800 audio frames on 24 NVIDIA V100 32G GPUs, following the same training strategy as F5-TTS.
For inference, we set the DCFG strength $\alpha_{speech}$ and $\alpha_{env}$ to 2.0.
Audio samples can be accessed at \href{https://yxlu-0102.github.io/DAIEN-TTS}{https://yxlu-0102.github.io/DAIEN-TTS}.

\vspace{-1mm}
\begin{table*}[t!]\scriptsize
  \caption{Subjective and objective evaluation results of the environment-aware TTS on the Seed-TTS \textit{test-en} set.}
  \label{tab: exp}
  \centering
  \resizebox{0.8\textwidth}{!}{
  \begin{tabular}{c|ccccc}
    \toprule
    Model  & WER(\%) $\downarrow$ & SIM-o $\uparrow$ & MOS $\uparrow$   & SSMOS $\uparrow$  & ESMOS $\uparrow$  \\
    \midrule
    \multicolumn{6}{c}{Silence Environment Prompt} \\
    \midrule
    Human   & 2.14 & 0.73 & 3.91 ($\pm$ 0.09) & 3.72 ($\pm$ 0.09) & - \\
    Vocoder & 2.18 & 0.70 & - & - & - \\
    F5-TTS (Clean Spk. Prompt)  & 2.30 & 0.58 & 3.80 ($\pm$ 0.09) & 3.60 ($\pm$ 0.09) & - \\
    \midrule
    F5-TTS  & 2.87 & 0.49 & 3.09 ($\pm$ 0.11) & 2.92 ($\pm$ 0.11) & - \\
    DAIEN-TTS (w/o CA) & 2.03 & 0.59 & 3.81 ($\pm$ 0.08) & 3.60 ($\pm$ 0.09) & - \\
    DAIEN-TTS & \textbf{1.93} & \textbf{0.60} & \textbf{3.84 ($\pm$ 0.09)} & \textbf{3.64 ($\pm$ 0.09)} & - \\
    \midrule
    \multicolumn{6}{c}{Background Environment Prompt} \\
    \midrule
    Human + Environment  & 2.80 & 0.70 & 3.86 ($\pm$ 0.08) & 3.81 ($\pm$ 0.08) & 3.72 ($\pm$ 0.08) \\
    Vocoder   & 3.03 & 0.65 & - & - & - \\
    \midrule
    DAIEN-TTS (w/o CA) & 2.93 & 0.54 & 3.68 ($\pm$ 0.10) & 3.70 ($\pm$ 0.09) & 3.49 ($\pm$ 0.10) \\
    DAIEN-TTS & \textbf{2.83} & \textbf{0.55} & \textbf{3.78 ($\pm$ 0.08)} & \textbf{3.73 ($\pm$ 0.08)} & \textbf{3.65 ($\pm$ 0.08)} \\
    \bottomrule
  \end{tabular}}
  \vspace{-2mm}
\end{table*}
\vspace{-1mm}
\subsection{Baselines and Evaluation Metrics}
\vspace{-1mm}
Since the core of environment-aware TTS lies in the disentanglement and reconstruction of the background environment, we evaluate our proposed DAIEN-TTS under two scenarios:
1) We first use silence as the environment prompt to synthesize clean personalized speech, in order to evaluate the model’s ability to disentangle the environment component from the speech component.
For baseline systems, we employed F5-TTS and trained it on the LibriTTS dataset for 600k steps.
Furthermore, we implemented an ablated version of DAIEN-TTS, referred to as DAIEN-TTS (w/o CA), in which the cross-attention layers in the DiT blocks are removed, and the environment, speaker, and text conditions are directly concatenated, following UmbraTTS.
2) Based on the first scenario, we further used background environment prompts to evaluate the model’s ability to reconstruct the environment component in the synthesized environmental speech.
Since there were no existing baselines for time-varying environment-aware TTS, we solely compared DAIEN-TTS with its ablated variant DAIEN-TTS (w/o CA).

We evaluated the proposed DAIEN-TTS and the baselines using both objective and subjective metrics.
For the objective evaluation, speech intelligibility was measured by computing the word error rate (WER) using transcriptions generated by Whisper-large-v3 \cite{radford2023robust}.
Speaker similarity (SIM-o) was quantified by calculating the cosine similarity between speaker embeddings extracted from the synthesized speech and the clean speaker prompt, using the WavLM-large-based speaker verification model \cite{chen2022large}.
For the subjective evaluation, we conducted three mean opinion score (MOS) tests: MOS for naturalness, speaker similarity MOS (SSMOS), and environment similarity MOS (ESMOS), assessing the perceived naturalness, speaker similarity, and environment similarity, respectively.
We crowd-sourced more than 30 raters on Amazon Mechanical Turk, with each rater scoring 20 audio samples on a 5-point scale with 0.5-point resolution.

\vspace{-1mm}
\section{Results and Analysis}
\vspace{-1mm}
\label{sec:res}
\subsection{Evaluation with Silence Environment Prompts}
The results of environment-aware TTS with silence environment prompts are shown in the upper part of Table~\ref{tab: exp}. 
We first took the human recordings from the Seed-TTS \textit{test-en} set as the ground truth (denoted as ``Human''), and resynthesized them with the vocoder (denoted as ``Vocoder'').
Experimental results demonstrated that the vocoder resynthesized ground truth exhibited slight degradations in SIM-o and WER, which can serve as the upper bound for clean synthesized speech in terms of objective metrics.
For the F5-TTS baseline trained on clean speech data, the synthesized speech exhibited high naturalness, intelligibility, and speaker similarity when provided clean speaker prompts.
However, a noticeable drop in both objective and subjective metrics was observed when environmental speaker prompts were conditioned, indicating that F5-TTS was sensitive to background environment.

In contrast, the proposed DAIEN-TTS and its variant significantly outperformed F5-TTS under environmental speaker prompts across all the metrics.
These results suggested that the SES module effectively disentangled the background environment from the environmental speaker prompt while maintaining high fidelity in the separated speech component.
Moreover, the DAIEN-TTS variants even slightly surpassed F5-TTS with clean speaker prompts in all metrics, which may be attributed to the data augmentation effect introduced by training on paired clean–environmental speech data.
Finally, within the DAIEN-TTS variants, comparable performance was observed, suggesting that removing the cross-attention layer did not significantly impact synthesis quality when the environment condition is silent.
This further implied that the cross-attention layers primarily contributed to modeling the background environment.

\subsection{Evaluation with Background Environment Prompts}
The results of the environment-aware TTS with background environment prompts are presented in the bottom part of Table~\ref{tab: exp}.
To construct ground truth, we mixed the human recordings with the background environment audio according to the SNR of the environment prompt (denoted as ``Human + Environment''), and resynthesized them using a vocoder retrained on the environmental speech data, serving as the upper bound of the synthesized environmental speech.
In this scenario, the DAIEN-TTS variants still surpassed the vocoder-resynthesized ground truth in terms of WER, and attained competitive SIM-o scores (approximately 0.1 lower than ``Vocoder'').
Moreover, our proposed DAIEN-TTS attained MOS, SSMOS, and ESMOS scores comparable to the human recordings, highlighting its effectiveness in environment reconstruction.

Between the variants, the proposed DAIEN-TTS slightly outperformed DAIEN-TTS (w/o CA) in speaker similarity, as reflected by SIM-o and SSMOS, while demonstrating superior naturalness and environmental fidelity, as indicated by MOS and ESMOS.
These results indicated that incorporating environmental information via cross-attention is more effective for modeling background environments than the direct concatenation approach used in DAIEN-TTS (w/o CA) and UmbraTTS. 
The distortion in the reconstructed background environment by DAIEN-TTS (w/o CA) further impacted the naturalness of the synthesized speech.
Overall, the evaluations across both scenarios demonstrated that our proposed DAIEN-TTS successfully disentangled and reconstructed the speaker and background environment components, achieving high-quality environment-aware TTS.

\vspace{-2mm}
\section{Conclusions}
\vspace{-2mm}
\label{sec:conclude}
In this paper, we introduced DAIEN-TTS, an environment-aware zero-shot TTS method that enabled independent control of the timbre and time-varying background environment in synthesized speech via disentangled audio infilling.
By leveraging a pretrained SES module for speech–environment separation and a flow-matching-based TTS framework for environmental mel-spectrogram infilling, our approach effectively disentangled and reconstructed environmental speech components.
During inference, the DCFG mechanism was employed to enhance the controllability across speech and environment components, while the SNR adaptation method ensured consistency between the synthesized environmental speech and the environment prompt.
Experimental results demonstrated the superiority of the proposed DAIEN-TTS in terms of naturalness, timbre control, and environment control of the synthesized speech.
In future work, we will explore the mutual influence between speech and the background environment.
\vfill\pagebreak

\bibliographystyle{IEEEbib}
\bibliography{refs}

\end{document}